# Optical anapole mode in nanostructured lithium niobate for enhancing second harmonic generation

Yang Li[a], Zhijin Huang[a], Zhan Sui, Huajiang Chen, Xinyue Zhang, Heyuan Guan, Wentao Qiu*, Jiangli Dong, Wenguo Zhu, Jianhui Yu, Huihui Lu*, Zhe Chen

[a] **Yang Li and Zhijin Huang:** These authors equally contributed to this work

**\*Corresponding authors: Wentao Qiu**, Key Laboratory of Optoelectronic Information and Sensing Technologies of Guangdong Higher Education Institutes, Jinan University, Guangzhou 510632, China, e-mail: qiuwentao16@163.com; **Huihui Lu**, Guangdong Provincial Key Laboratory of Optical Fiber Sensing and Communications, Jinan University, Guangzhou 510632, P.R. China, e-mail: thuihuilu@jnu.edu.cn

**Yang Li, Zhijin Huang, Huajiang Chen, Xinyue Zhang, Heyuan Guan, Jiangli Dong, Wenguo Zhu, Jianhui Yu, Zhe Chen:** Guangdong Provincial Key Laboratory of Optical Fiber Sensing and Communications, Jinan University, Guangzhou 510632, P.R. China

**Zhan Sui:** Shanghai Institute of Laser Plasma, China Academy of Engineering Physics, Shanghai, 201800, China

**Abstract:** Second harmonic generation (SHG) with a material of large transparency is an attractive way of generating coherent light sources at exotic wavelength range such as VUV, UV and visible light. It is of critical importance to improve nonlinear conversion efficiency in order to find practical applications in quantum light source and high resolution nonlinear microscopy, etc. Here an enhanced SHG with conversion efficiency up to $10^{-2}$ % at SH wavelength of 282 nm under 11 GW/cm$^2$ pump power via the excitation of anapole in lithium niobite (LiNbO$_3$, or LN) nanodisk through the dominating d$_{33}$ nonlinear coefficient is investigated. The anapole has advantages of strongly suppressing far-field scattering and well-confined internal field which helps to boost the nonlinear conversion. Anapoles in LN nanodisk is facilitated by high index contrast between LN and substrate with properties of near-zero-index via hyperbolic metamaterial structure design. By tailoring the multi-layers structure of hyperbolic metamaterials, the anapole excitation wavelength can be tuned at different wavelengths. It indicates that an enhanced SHG can be achieved at a wide range of pump light wavelengths via different design of the epsilon-near-zero (ENZ) hyperbolic metamaterials substrates. The proposed nanostructure in this work might hold significances for the enhanced light-matter interactions at the nanoscale such as integrated optics.

**Keywords:** anapole; hyperbolic metamaterials; lithium niobate; second harmonic generation (SHG).

## 1 Introduction

Second harmonic generation (SHG) is a nonlinear optical process which combines two photons at the fundamental frequency (FF) and convert them into one photon at the second harmonic wavelength (SH) [1]. SHG has a wealth of applications in such as coherent laser sources [2], biological imaging microscopy [3], etc. The key of a practical SHG device is its conversion efficiency. Confining light in nonlinear material is a way of boosting nonlinear response.

In recent years, there is a growing interest in the study of generating anapoles for various applications due to its strong light confinement ability [4, 5]. An anapole occurs when the far field radiation by electric dipole (ED) and toroidal dipole (TD) modes in Cartesian coordinates cancel each other out, and it is a non-radiative mode characterized by suppressing far field radiation therefore yielding a strong near field confinement [6, 7]. An efficient nonlinear device is feasible combining strong light confinement provided by anapoles and materials with high nonlinearity. As an example, the THG process has been greatly enhanced on germanium nanodisk via the excitation of anapole mode with near-infrared pump light [8]. And an anapole mode facilitate the generation of SH in LN with aluminum substrate yielding a high conversion efficiency of $1.1528 \times 10^{-3}\%$ at VUV regime [9].

However, these anapoles are generated in nonlinear materials with high refractive index and a relatively narrow transparency range. In terms of wideband transparency and large nonlinearity, LN is a good choice, since it has remarkable physical properties such as electro-optic, piezoelectric, pyroelectric and SHG effect [1, 10]. It is very stable crystal with trigonal symmetry and a large transparency window. LN has a large second order nonlinearity coefficient $d_{33}$ of 25 pm/V which is oftentimes exploited as SHG devices by periodically poled lithium niobite (PPLN) configurations [11, 12]. A relatively low refractive index (extraordinary refractive index $n_e$=2.25 at 500 nm) of LN than those such as silicon [13, 14] or germanium [8, 15] makes it rarely exploited in anapole configuration since high index contrast between substrate and material is important.

In order to enhance the index contrast between LN nanostructure and substrate, metamaterials which can be tuned to be as ENZ or even negative refractive index are chosen. Among these metamaterials, hyperbolic metamaterials (HMMs), as the topology of the isofrequency surface are hyperbolic curves, have been widely investigated recently [16]. In general, HMMs can be achieved by consisting of alternating layers of metal and dielectric [17] or as an array of nanowires [16]. Here HMMs are employed here exploiting their near ENZ and low effective permittivity property which assists a high index contrast between LN nanostructure and the substrate.

In this letter, we theoretically present a nanostructure for SHG application as shown in Figure 1. The nanostructure consists of a LN-nanodisk on the HMM substrate with a silica ($SiO_2$) spacer layer. A high normalized electric field intensity enhancement up to ~35 at the resonant anapole is achieved. The SHG conversion efficiency with no need of phase-matching condition yields as high as $5.1371 \times 10^{-5}$ thanks to the strong local field enhancement via anapole excitation.

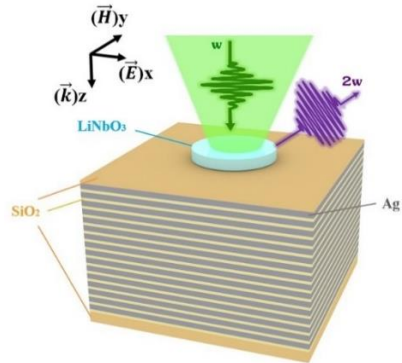

**Figure 1:** Schematic depiction of the proposed nanostructure for generating SH. The nanostructure consists of three parts: LN nanodisk on the top, HMM at the bottom consisting of 16 bilayers of alternating silver (Ag) [18] and silicon dioxide ($SiO_2$) [10] layers and a $SiO_2$ spacer layer with particular designed thickness sandwiched between LN nanodisk and HMM. The nanostructure lies on a glass substrate. The vector $\vec{k}$ represents the wave vector of the incident light and $\vec{E}$ indicates the x-polarized incident field.

The paper is structured as follows: firstly, the dispersion property of the multilayered HMM is engineered and geometry parameters of nanostructure are optimized. Secondly, the scattering spectra and electric-field distributions are investigated to achieve the anapole resonance, which is further analyzed by the multipolar decomposition method. Afterwards, the SHG performance is evaluated by nonlinear simulation of the optimal nanostructure. The dependences of SHG conversion efficiency with respect to the parameters of the input source are investigated. At last, comparisons of the state of the art in SHG are presented.

## 2 Optimization of HMM

In order to achieve high index contrast between LN nanodisk and substrate facilitating anapole excitation, a low effective refractive index substrate constituted of HMM is employed. In general, HMMs are made of alternating subwavelength-thickness layers of metal and dielectric which have opposite sign of permittivity facilitating a large tuning range of effective permittivity. The effective permittivity of HMMs are oftentimes investigated via effective medium theory (EMT) [16, 19]. In short, EMT allows to estimate the parallel component of effective permittivity $\varepsilon_\parallel < 0$ and perpendicular component of effective permittivity $\varepsilon_\perp > 0$ of HMMs which is related to filling fraction and the permittivity of the constituent materials.

The filling fraction of the metal ($\rho$) is defined as follow [16]:

$$\rho = \frac{d_{metal}}{d_{metal} + d_{dielectric}} \quad (1)$$

Where $d_{metal}$ and $d_{dielectric}$ are the thicknesses of the metallic and the dielectric layers respectively. The equations for estimating the effective permittivity of the parallel ($\varepsilon_{xx} = \varepsilon_{yy} = \varepsilon_\parallel$) and perpendicular ($\varepsilon_{zz} = \varepsilon_\perp$) components based on field boundary conditions in Maxwell's Equations are as follows [20]:

$$\varepsilon_\parallel = \rho \varepsilon_{metal} + (1-\rho) \varepsilon_{dielectric} \quad (2)$$

$$\varepsilon_\perp = \frac{\varepsilon_{metal} \varepsilon_{dielectric}}{\rho \varepsilon_{dielectric} + (1-\rho) \varepsilon_{metal}} \quad (3)$$

Where $\varepsilon_{metal}$ and $\varepsilon_{dielectric}$ are the permittivities of the metallic and the dielectric layers respectively. According to the sign of $\varepsilon_\parallel$ and $\varepsilon_\perp$, HMMs are classifies as Type I and Type II. Type I corresponds to a positive $\varepsilon_\parallel$ and negative $\varepsilon_\perp$. Whereas, Type II is reversed. Since the parallel index contrast is a key in the excitation of anapoles, a low or even negative $\varepsilon_\parallel$ corresponding to Type II HMMs is chosen. On the other hand, the Type II HMMs with uniaxial anisotropic ($\varepsilon_{xx}, \varepsilon_{yy} < 0; \varepsilon_{zz} > 0$) show more metallic characteristics such as low transmission and high reflection which is beneficial for supporting the anapole excitation inside LN nanodisk [21, 22].

The target wavelength of SHG is short wavelength around 200 nm-300 nm, therefore HMMs that allow a range of 200 nm-1000 nm covering the pump light wavelength and SH is favorable. In terms of metallic materials, silver（Ag）which has small absorption loss and small real part of the permittivity in the targeted wavelength is more suitable than some others such as gold (Au)[18] or aluminum (Al) [10]. With regarding to dielectric material, SiO$_2$ which has weak absorption and small permittivity in the targeted wavelength is chosen. On the other hand, the chosen material Ag and SiO$_2$ are oftentimes investigated in thick film layer configurations in the literature with mature film deposition technique guaranteeing the possibility of structure fabrications.

We can tailor the dispersion of HMMs by appropriate filling fraction of metal (Ag) and dielectric (SiO$_2$). In this study, HMM is chosen as 16 pairs of alternating layers of Ag and SiO$_2$ thin films, with thicknesses of 30 nm and 15 nm respectively. The total thickness of the bilayer consisting of Ag and SiO$_2$ is fixed at 45 nm since this thickness yields small transmission losses and the filling fraction of the metal ($\rho$) is set as 0.67 in order to cover the SH wavelength.

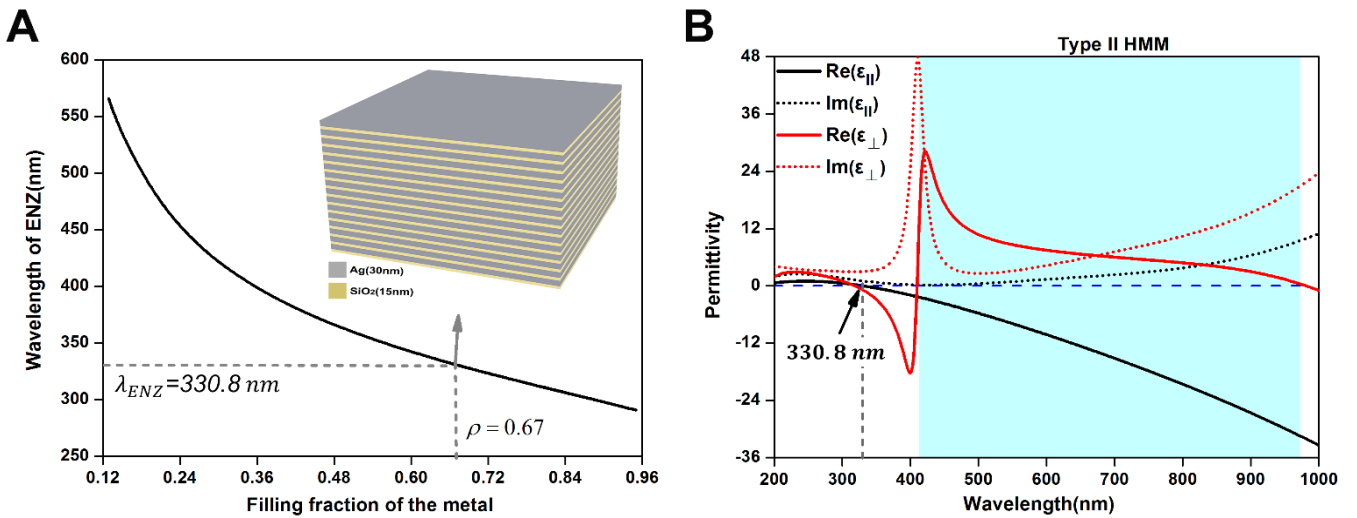

**Figure 2:** Optical properties of the HMM configuration.

(Color online) (A) The ENZ wavelength as a function of filling fraction of the metal (Ag) versus dielectric (SiO$_2$). The inset is the schematic of the Ag/SiO$_2$ multilayered HMM with $\rho = 0.67$, corresponding Ag thickness T(Ag)= 30 nm and SiO$_2$ thickness T(SiO$_2$)= 15 nm, where ENZ occurs at around 330.8 nm. (B) Real and imaginary parts of the parallel and perpendicular component of effective permittivity as functions of wavelength for Ag/ SiO$_2$ HMM with multilayered structures with filling fraction of the metal versus SiO$_2$ $\rho = 0.67$. The blue area corresponds to Type II HMM where parallel component of effective permittivity $\varepsilon_\parallel < 0$ and perpendicular component of effective permittivity $\varepsilon_\perp > 0$.

The ENZ wavelength versus the filling fraction of the metal ($\rho$) is investigated by EMT and it is depicted in Figure 2A. Taking into account the feasible film deposition thickness and the operation wavelength regime, a filling fraction of ($\rho$) is chosen as 0.67 (shown as dashed line in Figure 2A), which yields Ag thickness T(Ag)= 30 nm and SiO$_2$ thickness T(SiO$_2$)= 15 nm where the period is fixed at 45 nm. The ENZ wavelength occurs at around 330.8 nm. In terms of the total number of alternating thin film pairs, a number of 16 pairs is chosen to provide optimal performance since it yields a good compromise between the fabrication feasibility and small transmission losses.

The parallel and perpendicular component of effective permittivity corresponding to 16 pairs of alternating layers of Ag and SiO$_2$ thin films (30 nm Ag and 15 nm SiO$_2$) as functions of wavelength are shown in Figure 2B where the blue area corresponds to Type II HMM regime. This HMM configuration will be employed as a low effective index substrate supporting LN nanodisk whereas the material and thickness of terminating layer i.e. interfacing between HMM and LN nanodisk will be tuned according to its ability of high field enhancement which will be presented in the next section.

## 3 Results and discussions

### 3.1 Optimization of LN nanostructure

The purpose to optimize the nanostructured LN-nanodisk geometry is to obtain high field enhancement in nonlinear material therefore enhancing nonlinear response. In order to achieve that, the spacer layer between LN nanodisk and HMMs is SiO$_2$ since it has negligible absorption loss within the considered wavelength range. The optimal thickness of SiO$_2$ spacer layer and geometries of LN nanodisk are determined via three-dimensional finite-difference time-domain (Lumerical FDTD) simulations. The incident field is a broad bandwidth total-field scattered-field (TFSF) source and the simulated structure is shown in Figure 1. Perfectly matched layer (PML) boundary conditions are employed in all three dimensions simulating a finite structure size. The optimal SiO$_2$ spacer layer thickness is the one that allows high field enhancement in LN nanostructure [23, 24]. Therefore, the averaged intensity enhancement which is defined as the ratio of the averaged volumetric field integration in LN nanoparticle with that of field amplitude $E_0$ in the incident wave as $\langle |E/E_0|^2 \rangle$.

Figure 3A shows the calculated maximum averaged intensity enhancement when varying the spacer thickness T whereas the LN nanodisk is centered on top with parameters of diameter (D=432 nm) and height (H=104 nm). The maximum averaged intensity enhancement $\langle |E/E_0|^2 \rangle$ achieves at 21nm-thickness SiO$_2$ layer. The averaged intensity enhancement of the optimal SiO$_2$ spacer layer (black curve) and 0nm-thickness SiO$_2$ spacer layer (red curve) are calculated wavelength by wavelength with monochromatic wave excitation as shown in Figure 3B. The highest field enhancement occurs around 565.4 nm and it yields 1.3 times field enhancement compared with 0nm-thickness SiO$_2$ spacer layer configurations.

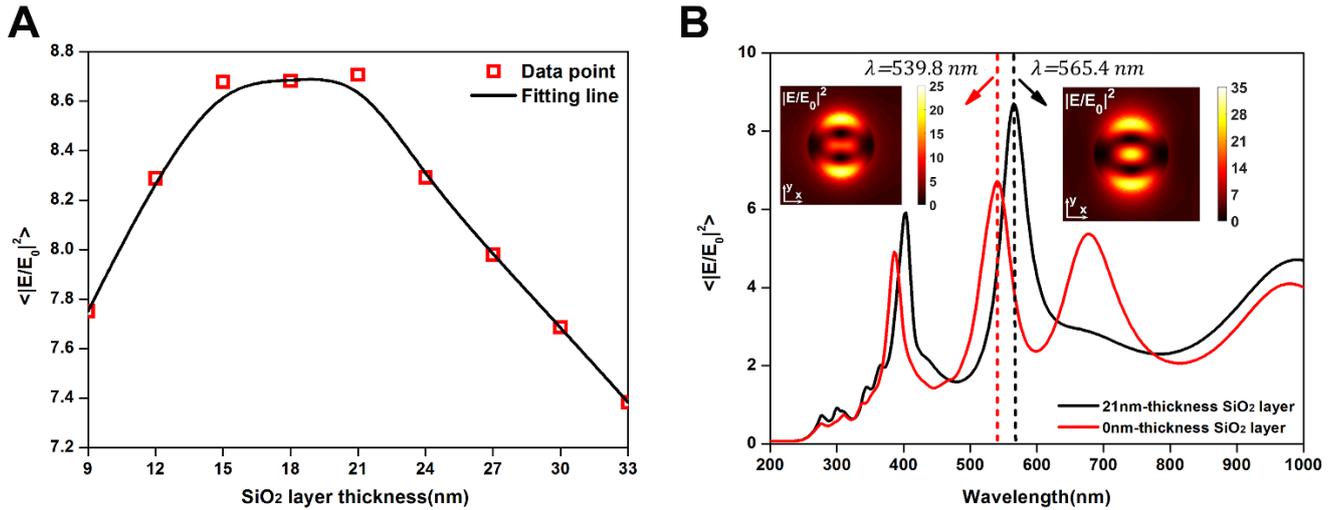

**Figure 3:** Optimization of the spacer-layer thickness.

(Color online) (A) The calculated maximum averaged electric field intensity enhancement as a function of $SiO_2$ spacer thickness T whereas the LN nanodisk is centered on top with parameters of diameter (D=432 nm) and height (H=104 nm). The data points are fitted. (B) Averaged intensity enhancement versus wavelength of 21nm-thickness $SiO_2$ layer (black curve) and 0nm-thickness $SiO_2$ layer (red curve). The insets show the electric field distributions at the highest field enhancement mode of each configuration respectively.

In terms of the geometries of LN nanostructure, LN (nanodisk) is the most efficient scatter comparing to other shapes such as cube or sphere[25]. A highly efficient scattering together with high index contrast are promising to the excitation of anapole which yields large field enhancement. Three-dimensional FDTD numerical simulations are employed to analyze different height and diameter of the LN nanodisk whereas the averaged field enhancement is calculated. The results are shown in Figure 4. Figure 4A and 4B show the normalized maximum electric field intensity enhancement versus varying one parameter while the other one is fixed and they indicate that optimal might lie in the vicinity of the diameter (D) of 432 nm and the height (H) of 104 nm.

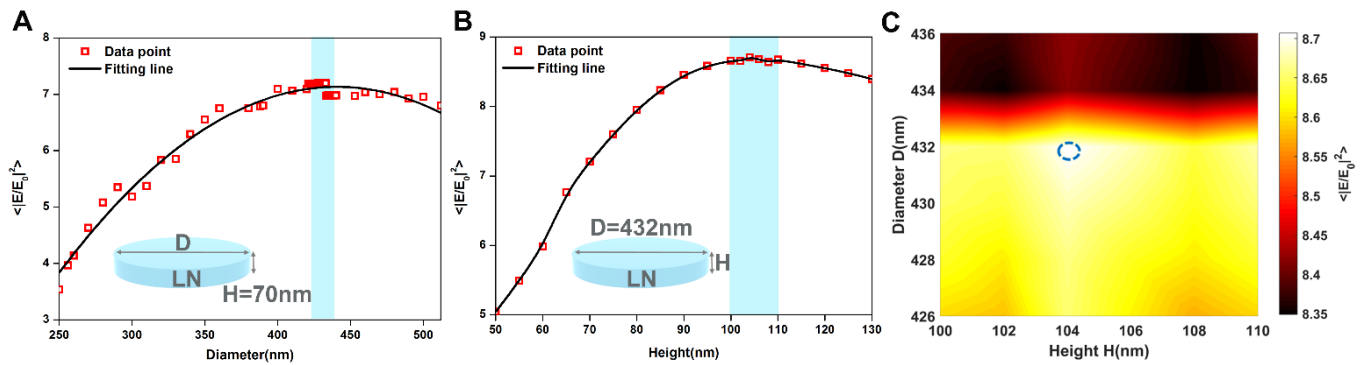

**Figure 4:** Numerical analysis of the LN-nanodisk geometries.

(Color online) (A) Normalized maximum electric field intensity enhancement of varying the height of LN nanodisk, while diameter is fixed to 432 nm. The coarse optimal D is around 432 nm. (B) Normalized electric field intensity enhancement of varying the diameter of LN nanodisk, while height is fixed to 70 nm. The coarse optimal H is around 104 nm. The data points are fitted. (C) Fine optimization of LN nanodisk geometries around the vicinity of coarse optimal one (D=432 nm, H= 104 nm) and the optimal one is depicted by the dashed blue circle corresponding to parameters of D=432 nm, H= 104 nm.

Fine optimization of LN nanodisk geometries around the vicinity of coarse optimal one (D=432 nm, H= 104 nm) is performed and the result is shown in Figure 4C. The optimal one is depicted by the dashed blue circle corresponding to parameters of diameter (D=432 nm) and height (H=104nm) which yields high field enhancement. A suitable height-to-diameter ratio (Height/Diameter defined as the geometrical aspect ratio) is also important for the potential of anapole excitations [13, 26] and here the optimal aspect ratio Height/Diameter is ~0.24.

Total scattering spectra of the optimal nanostructure are simulated and the averaged field enhancement factor in the wavelength range between 200 nm and 1000 nm is investigated as shown in Figure 5A.

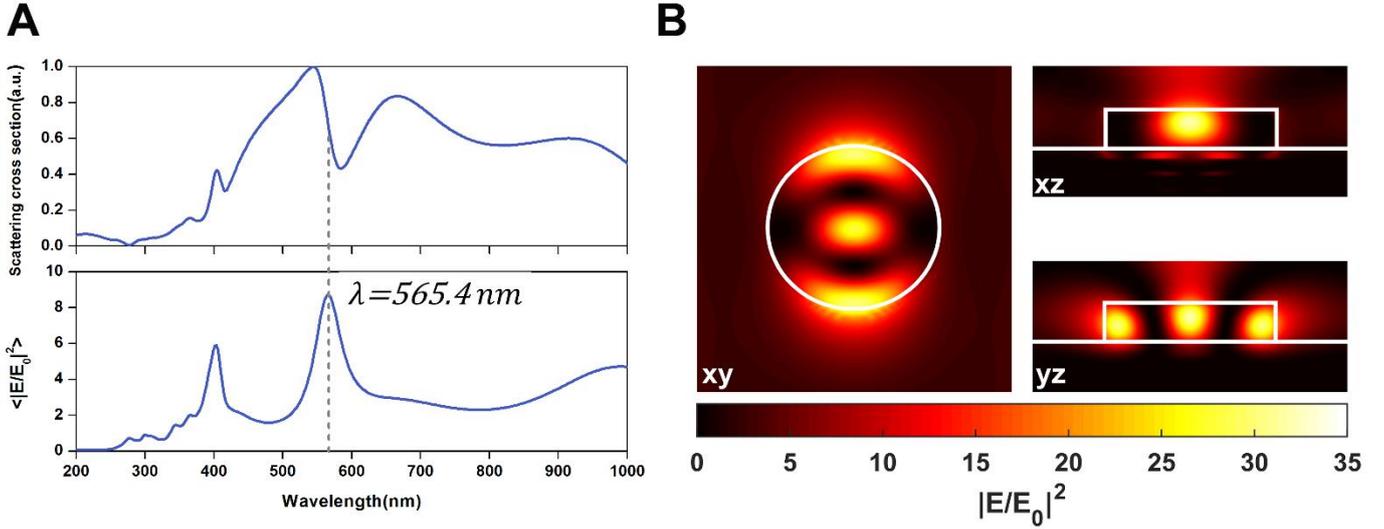

**Figure 5:** Simulation scattering spectra and field distributions of the LN nanostructure.
(Color online) (A) The top figure corresponds to the normalized scattering cross section and the bottom corresponds the intensity enhancement $\langle |E/E_0|^2 \rangle$ of the optimal configurations. The gray dashed line shows the mode of the highest field enhancement. (B) The normalized electrical field intensity enhancement distributions $|E/E_0|^2$ of LN nanodisk at the incident wavelength of 565.4 nm. The distributions of the normalized electric field in xy-, xz-, and yz-planes correspond to the dashed cured outlined mode in Figure 5A where the LN nanodisk (height of 104 nm and a diameter of 432 nm) is placed at the center with 21 nm SiO$_2$ spacer layer and 16 pairs of HMM sitting on glass substrate as the simulated nanostructure presented in Figure 1. The nanostructure of xz and yz cross sections are outline with white solid lines.

The top figure of Figure 5A corresponds normalized scattering cross section where a pronounced dip indicating suppression of far field radiation is observed in the scattering cross-section spectrum. The bottom figure of Figure 5A corresponds the averaged intensity enhancement $\langle |E/E_0|^2 \rangle$ of the optimal configurations where a $\langle |E/E_0|^2 \rangle$ of 8.7 enhancement is obtained which is almost an order-of-magnitude higher than the other reported nanostructures [9] thanks to the high index contrast between LN nanodisk and HMM substrate. The largest average field enhancement $\langle |E/E_0|^2 \rangle$ occurs at a wavelength of 565.4 nm which locates at the decreasing slop near the dip wavelength of the total scattering cross section.

Figure 5B shows the normalized electrical field intensity enhancement distributions in xy-, xz-, and yz-planes corresponding to the dashed cured outlined mode in Figure 5A. The maximum value of the normalized electric field intensity enhancement distributions $|E/E_0|^2$ can reach up to ~35 which is a high value compared to those reported in references [9, 14, 24] and from which one can see that great enhancement of electric field is confined inside the LN nanodisk. With this large field enhancement obtained and the large index contrast between LN nanodisk and HMMs, we will investigate whether this mode is an anapole or not via multipolar decomposition in the next section.

### 3.2 Multipolar decomposition

In order to understand the origin of the large obtained field enhancement at the local dip wavelength of the total scattering cross section, multipolar decomposition analysis is performed. The total scattering cross section can be decomposed as the sum of radiations from different electric and magnetic multipoles as follows [27, 28]:

$$C_{sca} = \frac{\pi}{k^2}\sum_{l=1}^{\infty}\sum_{m=-l}^{l}(2l+1)\left[\left|a_E(l,m)\right|^2 + \left|a_M(l,m)\right|^2\right] \quad (4)$$

Where $k$ is the incident wave vector, $l$ is the order of the multipole and $m$ is the amount of the z-component of angular momentum carried per photon.

In view of each multipole mode of multipole decomposition influenced by the nanostructural properties including the effects of spacer layer and complicated substrate on spectral shifts [26, 29, 30], the decomposition up to quadrupoles while neglecting other higher order multipoles is performed by finite element method (COMSOL Multiphysics) in spherical coordinates for the stability of calculations and the result is shown in Figure 6A. Due to the neglect of other higher order multipoles' contributions, the sum scattering cross section is a bit different from that displayed in the top figure of Figure 5A where all the multipoles contributions are taken into account [31, 32]. In general, different local dips in the sum scattering cross section correspond to anapoles of different order with different purity which are related to the degree of far field radiation extinction by destructive interference between Cartesian ED and TD modes. As such, the contributions from each multipole term can be clearly identified. Whereas at the local dip wavelength, spherical ED mode takes up an important portion. The dashed circled out local dip in Figure 6A indicates the effectively suppressed far field radiation which is termed anapole leading to a significant drop of the sum scattering cross section. Upon analyzing multipole contributions to the light-matter interactions, one can note that nearby the local dip wavelength of the spherical ED mode, the spherical magnetic quadrupole (MQ) resonance contributes to the far field radiation, making the anapole mode pure incompletely. Meanwhile, the contributions from spherical magnetic dipole (MD) and electric quadrupole (EQ) modes are negligible.

The spherical ED mode of radiation suppression in spherical coordinates normally is caused by interference with other multipoles. In order to find out this, TD mode which is neglected in spherical coordinates [33] multipolar decomposition is calculated by integration of the currents within the nanostructure in Cartesian coordinates.

In Cartesian coordinates, the radiated moments of Cartesian ED and TD can be derived from the scattering current density $\vec{J}$ as follows [34, 35].

$$ED: \vec{P} = \frac{1}{-i\omega}\int \vec{J}d^3r \quad (5)$$

$$TD: \vec{T} = \frac{1}{10c}\int\left[\left(\vec{r}\cdot\vec{J}\right)\vec{r} - 2r^2\vec{J}\right]d^3r \quad (6)$$

Where c is the speed of light in vacuum, $\vec{P} = -ik\vec{T}$ represents the contributions of the electric dipole and toroidal dipole modes in current Cartesian coordinates [6]. Figure 6B shows the Cartesian ED and TD multipole contributions in Cartesian coordinates. The slight wavelength difference between the scattering dip of spherical ED mode and the crossing point of the Cartesian ED and TD contributions results from the material losses [36].

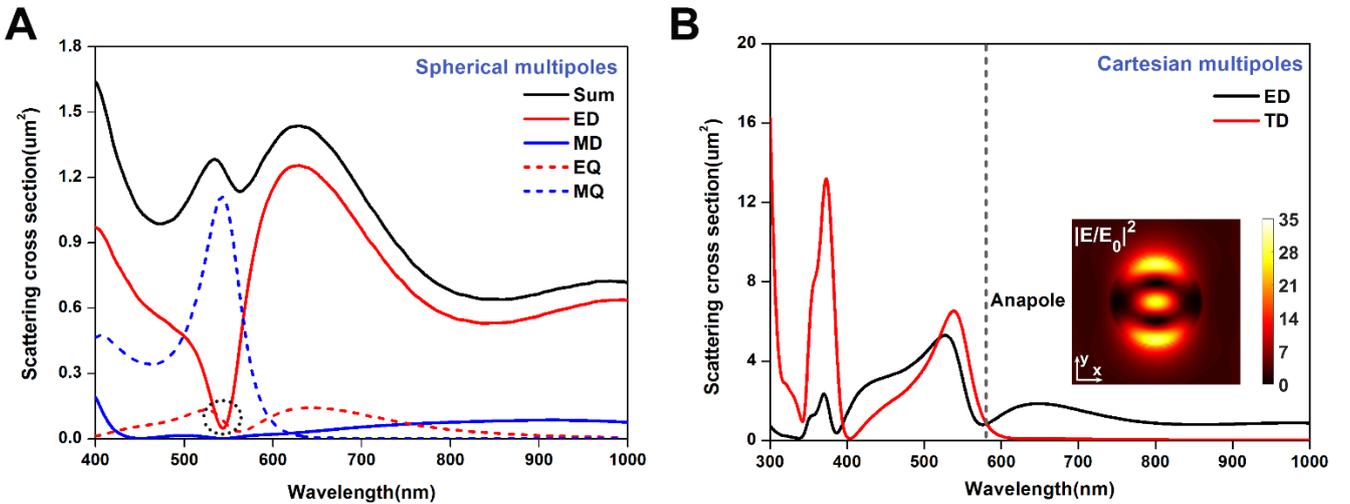

**Figure 6:** Multipolar analysis of light scattering by the nanostructure.
(Color online) (A) Multipoles decomposition of the studied nanostructure in spherical coordinates. Sum stands for sum scattering cross section from the decomposed multipoles up to the 2nd order, spherical ED for electric dipole, MD for magnetic dipole, EQ for electric

The intersection between the Cartesian ED and TD modes near 580.8 nm (gray dashed line in Figure 6B corresponding to the first-order anapole of this nanostructure) shows the destructively interference in the far field which yields to the excitation of anapole. The slight difference between the intersecting wavelength at 580.8 nm for far field and the resonant wavelength at 565.4 nm for near field (gray dashed line in Figure 5A) indicates the deviation of anapole wavelength resulting from the mismatch between the resonant positions in the near field and far field [37]. Therefore, both the suppression of spherical ED far field radiation and the intersection between the Cartesian ED and TD modes unambiguously reveal the physical origins of the first-order anapole mode of the nanostructure. The first-order anapole mode brings enhancement of the near field for realizing a large enhancement in SHG.

### 3.3 Nonlinear simulations

Thanks to the anapole excitation in the nanostructure, maximum field enhancement reaches up to ~35. SHG is studied employing the optimized parameters from above with LN nanodisk lying on HMMs and the substrate is SiO$_2$ (schematically shown in Figure 1). Phase-matching conditions [1] is not necessary due to the nanometrical thickness of the structure. In general, the nonlinear coefficient matrix of trigonal symmetric structure is as follows [1]:

$$d_{ijk} = \begin{bmatrix} 0 & 0 & 0 & 0 & d_{31} & -d_{22} \\ -d_{22} & d_{22} & 0 & d_{31} & 0 & 0 \\ d_{31} & d_{31} & d_{33} & 0 & 0 & 0 \end{bmatrix} \quad (7)$$

In LN the largest second order nonlinear susceptibility is $d_{33} = -41.7 \, pm/V$. Only the diagonal terms of second order nonlinear susceptibility will be considered in calculating nonlinear polarizations via FDTD Solutions. The peak pump intensity of the normal incident Gaussian pulse is set to $I_0 = 5.31 \, GW/cm^2$ (corresponding to the incident electric field amplitude of $1.9995 \times 10^8 \, V/m$) at 565.4 nm with pulses duration of 90 fs. It is a beam with a spot size one-micron wide in x-direction in the simulation estimating nonlinear response. The obtained power spectrum is shown in Figure 7A where there are two peaks corresponding to pump and SH respectively. As expected, the high fundamental field observed in Figure 7B induces an enhancement of the SHG field localized inside the LN nanodisk.

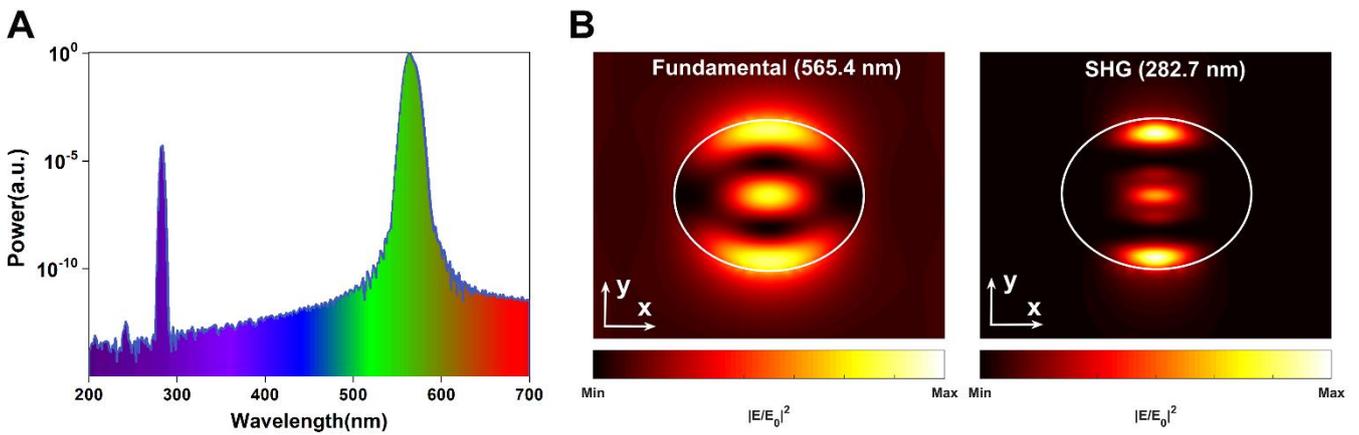

**Figure 7:** Nonlinear simulations of the nanostructure.
(Color online) (A) The normalized nonlinear power spectrum with spectral color where SH wavelength is at 282.7 nm and the fundamental wavelength is at 565.4 nm. (B) Normalized field intensity distributions in the xy planes at fundamental and SH wavelength. The white solid lines depict the xy cross sections of the nanostructure.

The SHG conversion efficiency is estimated via the equation as follows [38, 39]:

$$\eta_{SH} = \frac{P_{SH}}{P_{FF}} \quad (8)$$

$P_{FF}$ is the power of the fundamental wavelength i.e. the pump light, $P_{SH}$ is the power at the SH wavelength. Where $P = \frac{1}{2}\int real(\vec{S}) \cdot dA$, and $\vec{S}$ is Poynting vector $\vec{S} = \pm(\vec{E} \times \vec{H})/2$ and A is the area of nanostructure. The SHG conversion efficiency obtained in Figure 7A reaches up to $5.1371 \times 10^{-5}$ with pump light at 565.4 nm. It is revealed in Figure 7B that the strong enhancement of SHG conversion efficiency originated mainly from the localized concentrated electromagnetic field in the LN nanodisk induced by the anapole mode (i.e. at fundamental wavelength). The SHG conversion efficiency can be improved by optimizing the pump light conditions such as pulse duration and peak intensity. The results of the obtained SHG conversion efficiency with respect to pulse duration and peak intensity are shown in Figure 8 where the structure parameters are the same as that employed in Figure 7A. And the SHG power as a function of pump intensity is shown in the inset of Figure 8B.

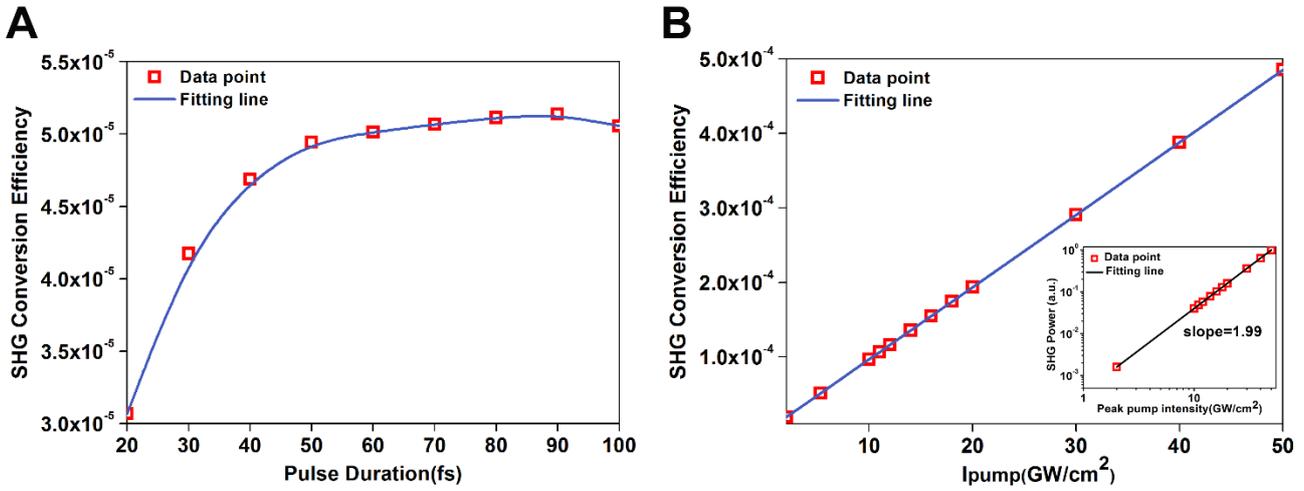

**Figure 8:** Optimizing calculations of the pump light conditions.
(Color online) (A) SHG conversion efficiency versus different pump light pulse duration. (B) SHG conversion efficiency as a function of pump light peak intensity varying from 2 GW/cm² to 50 GW/cm². The inset shows the log-log plot of quadratic dependence of the SHG power on the peak pump intensity at 565.4 nm fundamental wavelength. A fit to this data produces a black line with a slope of 1.99, which is very close to the theoretical prediction of 2 reasonably well.

As can be seen from Figure 8A, the optimal value of pulse duration is about 90fs which yields a SHG conversion efficiency up to $5.1371 \times 10^{-5}$. When the incident pulse duration is more than 90fs, the SHG conversion efficiency will saturate and decrease which is known as the cavity phenomena [40]. With regard to pump light peak intensity, the optical damage threshold of LN is considered as larger than 70 GW/cm² [1]. Therefore, the largest peak intensity in Figure 8B is considered only up to 50 GW/cm² and the largest obtained SHG conversion efficiency is at the order of ~$10^{-4}$. Also as shown in the inset of Figure 8B, the SHG power is proportional to the square of the pump intensity, which shows a very good agreement with the second-order nature of the SHG, confirming the SHG process.

As shown above in the article, the anapole-assisted nanostructure obtained here is to enhance nonlinear effects efficiently. Table 1 shows the comparisons of the optical field enhancement and SHG conversion efficiency of different structures where the proposed nanostructure in this article appears in bold. The SHG conversion efficiency is high up to the order of ~$10^{-4}$, which exceeds several orders higher than the case of other configurations such as nanoantennas and metasurfaces. Hence, the proposed nanostructure yielding higher conversion efficiency through anapole enhancement of the electromagnetic field might find applications in such as nanophotonics and biophotonics.

**Table 1:** Comparisons of the electric-field enhancement and SHG conversion efficiency of different structures.

| Structure | $\langle |E/E_0|^2 \rangle$ | $|E/E_0|^2$ | SHG conversion efficiency |
| --- | --- | --- | --- |
| Plasmonic and dielectric chiral nanostructure [41] | - | - | $\sim 10^{-11}$ |
| Nanocrystalline nanoparticle [42] | - | - | $1.5 \times 10^{-6}$ |
| Metal–dielectric hybrid nanoantennas [43] | - | 20 | $5 \times 10^{-6}$ |
| Broken symmetry Fano metasurfaces [44] | - | - | $\sim 6 \times 10^{-6}$ |
| LiNbO$_3$ nanodisks on an Al substrate [9] | ~6 | ~22 | $1.1528 \times 10^{-5}$ |
| Free-standing disk nanoantennas [45] | - | - | $1.9 \times 10^{-5}$ |
| GaAs based dielectric metasurfaces [46] | - | 30 | $\sim 2 \times 10^{-5}$ |
| **Nanostructured lithium niobite (this article)** | **8.7** | **~35** | **$5.1371 \times 10^{-5}$** |

## 4 Conclusion

In conclusion, an approach to enhance the near-field effects of nanostructure with higher index contrast between LN and HMM has been investigated in SHG numerically. The electric-field enhancement associated with optical anapole mode which occurs at the wavelength of 565.4 nm enhances efficiently nonlinear interactions where the achieved SHG conversion efficiency yields up to $10^{-4}$ under pump power of 11 GW/cm$^2$. The HMMs assisted anapole excitation in low refractive index material may also work for other materials in order to obtain an efficient nonlinear conversion. In such a configuration, the design of structure by tuning dispersion and optimization of geometric parameters for anapole excitation can also be applied to other nanostructures. The great flexibility for high tuning efficiency in these nanostructures might find some applications, such as spectrally selective device and compact optical device.

**Acknowledgments:** This work was supported by the National Natural Science Foundation of China (61775084, 61505069, 61705089, 61705087, 61675092), Guangdong Special Support Program (2016TQ03X962), Guangdong Natural Science Funds for Distinguish Young Scholar (2015A030306046), Science Foundation of Guangdong Province (2016A030310098, 2016A030311019), Fundamental Research Funds for the Central Universities (21619409, 21619410).